\documentclass{iau}

\usepackage{amsmath}
\usepackage{graphicx}
\usepackage{multirow}
\usepackage{orcidlink}
\begin{document}

\jnlPage{1}{7}
\jnlDoiYr{2025}
\doival{}

\aopheadtitle{Proceedings IAU Symposium}
\volno{398}
\editors{eds. Hyung Mok Lee, Rainer Spurzem and Jongsuk Hong}

\title{Black Holes in Globular Clusters: A Structural and Kinematic Perspective}

\author{Alessandro Della Croce~\orcidlink{https://orcid.org/0009-0005-9372-3885}}
\affiliation{Department of Astronomy, Indiana University, Swain West 727 E. 3rd Street, IN 47405 Bloomington, USA}

\begin{abstract}
Black holes (BHs) play a major role in the structural and dynamical evolutions of Globular Clusters (GCs).
Several recent works searched for BHs in Galactic GCs using scaling relations derived from numerical simulations. 
However, the conclusions drawn by such approaches are strongly dependent on the specific prescriptions adopted in numerical simulations.
Therefore, we analyzed a survey of 101 Monte Carlo simulations to identify some observable parameters that allow us to probe the present-day BH population in GCs reliably.
We thoroughly show that a single observable is not suited to infer the BH mass fraction in real GCs: similar values could be attained by systems with different BH mass fractions, depending on the specific dynamical evolution of the system.
Finally, we present a combination of observable parameters that efficiently breaks this degeneracy. We also compare values from numerical simulations with a sample of Galactic GCs.
\end{abstract}

\begin{keywords}
  methods: numerical
- stars: black holes
- stars: kinematics and dynamics
– globular clusters: general
- black hole physics
\end{keywords}

\maketitle

\section{Introduction}
The increasing number of detections of stellar mass BHs in Galactic GCs \citep[e.g.,][]{maccarrone_etal2007,Giesers_etal2018} challenged our previous view of the long-term retention of BHs in massive star clusters.
Recent studies have indeed shown that star clusters could retain a sizeable population of BHs for timescales longer than the Hubble time. 

Several works addressed the role of BHs in the long-term dynamical evolution of a stellar system. They showed that the presence of BHs, and more specifically, BH-BH binaries, acts as a source of energy for the system, thereby delaying its dynamical evolution \cite[e.g.,][]{breen_heggie2013}.
However, the early retention or ejection of BHs due to natal kicks and the distribution of kick velocities are still matters of intense investigation \citep[e.g.,][]{Belczynski_etal2002,janka_2013}.

The search for BHs in Galactic GCs addresses many fundamental and timely science cases: the early BH retention and natal kicks, the study of stellar dynamical interactions, 
and the BH-BH merging in dense stellar systems as a source of gravitational wave emission.
Recent studies tackled the inference of the total mass in stellar-mass BHs harbored by GCs.
In particular, \citet{askar_etal2018} 
used the luminosity density within the half-light radius as a probe for the BH subsystem properties.
On the other hand, \citet{weatherford_etal2020} used a theoretical correlation between the fraction of BHs and the degree of mass segregation to infer the present-day total mass in BHs within 50 Galactic GCs.

However, \citet[][see their Sect.~2.6]{askar_etal2018} pointed out that the inference of the mass in BHs using a single observable can be biased by specific assumptions adopted in the analysis.
Here, we thus address the degeneracies in the inference of the present-day BH population in GCs.
The purpose of this work is {\it i)} to point out that some previously adopted quantities are consistent both with systems with large BH mass fractions and with no (or a negligible fraction of) BHs; {\it ii)} to possibly identify the parameters that allow an unambiguous identification of the presence of BHs. 

In particular, we analyze a survey of 101 Monte Carlo simulations performed with the MOCCA code \citep{giersz_etal2013,hypki_etal2013} by investigating the system properties after 13~Gyr of evolution.
The simulation set was fully presented in \citet{bhat_etal2024}.
We highlight here that for each set of initial conditions, two simulations were performed assuming a different prescription for the BH natal kicks: either a broader distribution \citep[i.e., the same as neutron stars,][]{hobbs_etal2005}, or a reduced kick velocity based on the fallback prescription by \citet{Belczynski_etal2002}.

The results presented in this IAU proceedings are part of the work \cite{dellacroce_etal2024c}, and we refer to that paper for further details.

\section{Results}
\subsection{Numerical simulations}\label{sec:results_simulations}
For each simulation, we computed the total stellar mass ($M_{\rm GC}$) and the BH mass fraction (defined as $M_{\rm BH}/M_{\rm GC}$, with $M_{\rm BH}$ being the total mass in BHs).
We then explored the impact of a long-lived BH subsystem on the evolution of star clusters by studying several quantities: a parameter tracing the mass segregation state in the system (i.e., $\Delta$), the average luminosity density, the ratio between the core and the half-light radius ($R_{\rm hl}$), the cluster dynamical age, and the inverse of the equipartition mass.
All such quantities were found to be influenced by the presence of a massive BH population in the cluster center \citep[e.g.,][]{mackey_etal2008,bianchini_etal2016,askar_etal2018,weatherford_etal2020,aros_vesperini2023}

We recover the well-known role of BHs in delaying the dynamical evolution of the system driven by multiple two-body encounters: mass segregation, the collapse of the core, and evolution towards energy equipartition are all hampered by a massive system of BHs inhabiting the center of the cluster.
This process is reflected in a correlation between the BH mass fraction and the aforementioned quantities.
However, systems with much fewer (or even zero) BHs can show similar values at 13~Gyr.
These systems had long initial relaxation times and ejected BHs at formation due to large natal kicks.
The fact that clusters with very different BH mass fractions might exhibit similar properties highlights the pitfalls of inferring the BH mass fraction in real GCs using a single observable.

In this work, we introduce an observable parameter, which in synergy with $\Delta$ turns out to be particularly useful in discriminating between BH retention over a Hubble time and dynamical evolution. This is defined as the ratio between the 1D velocity dispersion ($\sigma_\mu$) computed within $0.2 R_{\rm hl}$ and at $R_{\rm hl}$ (hereafter $\sigma_\mu\,(<0.2R_{\rm hl}) / \sigma_\mu(R_{\rm hl})$).
This parameter probes the steepness of the velocity dispersion profile, which directly reflects the radial variation of the gravitational potential.

\begin{figure}[!h]
    \centering
    \includegraphics[width=0.5\linewidth]{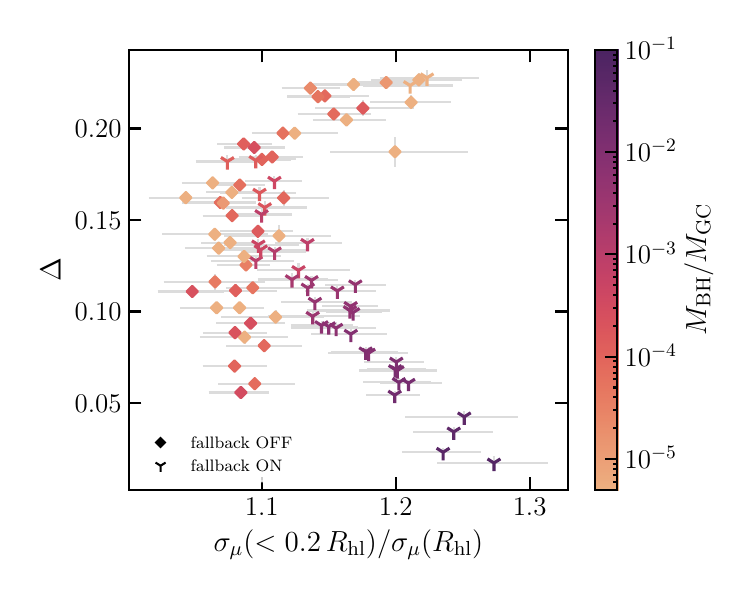}
    \caption{Mass segregation parameter as a function of the velocity dispersion ratio. Symbols show whether the simulation had (diamond) or did not have (upside-down triangle) the fallback prescription for BH formation. 
    Different colors show the BH mass fraction at 13~Gyr. The figure was reproduced from \citet{dellacroce_etal2024c}.}
    \label{fig:simulations}
\end{figure}

Figure~\ref{fig:simulations} shows $\Delta$ as a function of $\sigma_\mu\,(<0.2R_{\rm hl}) / \sigma_\mu(R_{\rm hl})$. Low mass segregation levels can be interpreted either as due to the presence of a massive BH subsystem (darker points) or due to the system being dynamically younger, without requiring high BH mass fractions (lighter points). 
However, these systems show different $\sigma_\mu\,(<0.2R_{\rm hl}) / \sigma_\mu(R_{\rm hl})$ values: the presence of a massive BH subsystem deepens the potential well in the central regions, increasing the velocity dispersion ratio. 
On the other hand, dynamically young systems without many BHs exhibit lower values.

\subsection{Observations}
\begin{figure}[!h]
    \centering
    \includegraphics[width=0.47\linewidth]{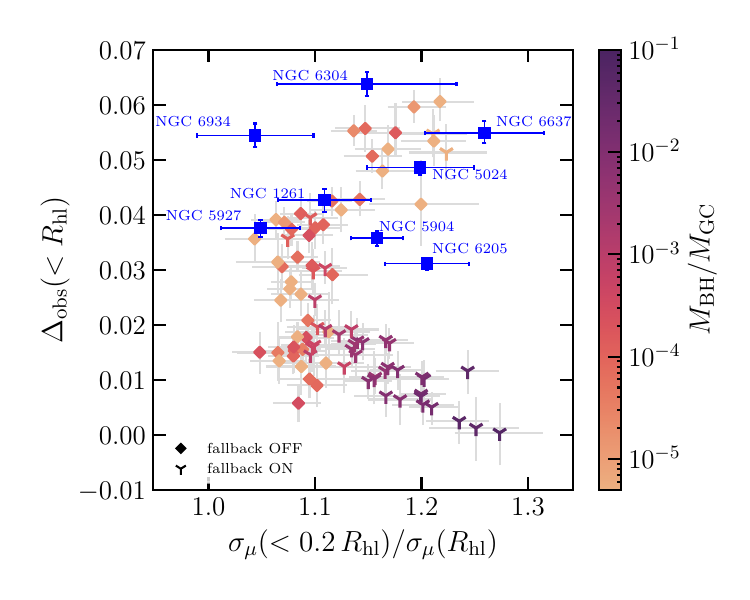}
    \includegraphics[width=0.47\linewidth]{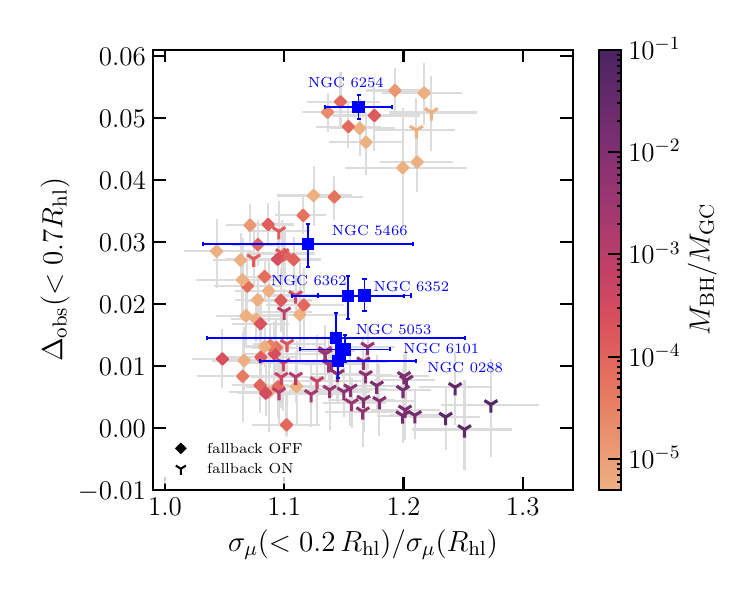}
    \caption{$\Delta_{\rm obs}$ within $R_{\rm hl}$ (left panel) and $0.7 R_{\rm hl}$ (right panel) as a function of the velocity dispersion ratio.
    Simulation values were recomputed by adopting the same magnitude and spatial selections as in the observations.
    In blue, we show the values (along with error bars) obtained for Galactic GCs with at least 100 stars with PM measurements. The figure was reproduced from \citet{dellacroce_etal2024c}.}
    \label{fig:observations}
\end{figure}

The parameters presented in Sect.~\ref{sec:results_simulations} can be measured in real GCs.
We thus selected Galactic GCs with both photometric data from \citet{sarajedini_etal2007} and individual star proper motion (PM) measurements by \citet{libralato_etal2022}, covering at least the central $0.7 R_{\rm hl}$. 
Such a selection included GCs previosly suggested to host a large fraction of stellar-mass BH (such as NGC~5053, NGC~6101, and NGC~6362, see e.g., \citealt{askar_etal2018,weatherford_etal2020}).

We used the PM catalogs to derive $\sigma_\mu$ using individual star measurements.
To quantify mass segregation, we used the photometric catalog after accounting for the spatial and magnitude-dependent completeness.
In particular, we defined the parameter $\Delta_{\rm obs}(<R_{\rm lim})$. Similarly to the $\Delta$ parameter shown in Fig.~\ref{fig:simulations}, this parameter traces the degree of segregation by quantifying the differences in the radial distributions between a brighter (i.e., more massive) and a fainter (i.e., lower in mass) population.
This is done up to a limiting radius $R_{\rm lim}$.

Figure~\ref{fig:observations} shows the simulations and observations comparison for clusters with different spatial coverages from observations. Values from numerical simulations were redefined following the selections adopted in the observations.
We first highlight here that decreasing the radial and mass ranges for the calculation of $\Delta$-like quantities almost hampers a proper distinction between systems with or without BHs, even when using numerical simulations.

Overall, we find a good agreement between simulation and observation values, except for a few systems (like NGC~6934 or NGC~6205). For these clusters, the simulation survey does not provide a good description of their present-day properties.

Some GCs exhibit little mass segregation, namely NGC~288, NGC~5053, and NGC~6101. This feature could be the imprint of a BH subsystem in these GCs or the result of a slower dynamical evolution.
We notice that while there might be hints of higher values in $\sigma_\mu\,(<0.2R_{\rm hl}) / \sigma_\mu(R_{\rm hl})$ which, combined with low mass segregation, would favor the former interpretation, observational errors do not allow us to discriminate between the possible scenarios fully.

\section{Conclusions}
We demonstrated that the role of BHs on the internal GC dynamics over their lifetime cannot be encapsulated in a single observable quantity. Indeed, clusters with different BH populations and different dynamical evolutionary histories may exhibit similar features when looked at using a single parameter. 
Therefore, multiple physical properties should be used to infer the present-day BH populations in real GCs.
As a promising candidate pair to do so, we propose $\Delta$ and $\sigma_\mu\,(<0.2R_{\rm hl}) / \sigma_\mu(R_{\rm hl})$, tracing both the structural and internal kinematics of the cluster.
Finally, we found that the current state-of-the-art data do not provide stringent enough constraints to fully discriminate between different scenarios.
Future astrometric and photometric data provided by, for instance, the \emph{Roman} space telescope may allow us to shed light on the subject.

\end{document}